\documentclass[final,sort&compress]{aipproc}            
\layoutstyle{6x9}

\begin{document}

\title{Partially Quenched Chiral Perturbation Theory to NNLO}

\classification{12.38.Gc, 11.30.Rd, 12.39.Fe}
\keywords{Chiral Perturbation Theory, Lattice QCD}

\author{Timo L\"ahde}{
  address={Helmholtz-Institut f\"ur Strahlen- und Kernphysik (HISKP),
           Bonn University, \\ Nu\ss allee 14-16, D-53115 Bonn, Germany}
}

\author{Johan Bijnens}{
  address={Department of Theoretical Physics, 
           Lund University, \\ S\"olvegatan 14A, SE-22362 Lund, Sweden}
}

\author{Niclas Danielsson}{
  address={Department of Theoretical Physics, 
           Lund University, \\ S\"olvegatan 14A, SE-22362 Lund, Sweden}
}

\begin{abstract}
This paper summarizes the recent calculations of the masses and 
decay constants of the pseudoscalar mesons at the two-loop level, or 
NNLO, in Partially Quenched Chiral Perturbation theory (PQ$\chi$PT). 
Possible applications include chiral extrapolations of Lattice QCD, as well 
as the determination of the low-energy constants (LEC:s) of QCD.
\end{abstract}

\maketitle

\vskip-10cm
\hspace{10cm}
\parbox{4cm}{\flushright{HISKP-TH 05/24\\LU-TP 05-48} \vskip8.2cm}

\section{Introduction}

Chiral Perturbation Theory ($\chi$PT) is an effective field theory 
approximation to QCD which allows for a description of low-energy hadronic 
properties in terms of a systematic expansion in momenta, energies and 
quark masses~\cite{ChPT}. $\chi$PT is by now a well developed field, such 
that many quantites, e.g. meson masses, decay constants and vacuum 
expectation values, are known up to NNLO, or $\mathcal{O}(p^6)$. As 
$\chi$PT is a non-renormalizable theory, an increasing, but certainly not 
overwhelming, number of low-energy constants (LEC:s) appears at each order 
in the calculations. These LEC:s are not determined by chiral symmetry, but 
are in principle calculable from full QCD. In addition, some of the LEC:s 
can also be fixed by fitting $\chi$PT expressions to experimental data.

At the present time, however, the only known way of deriving quantitative 
information on the low-energy properties of QCD from first principles is by 
means of Lattice QCD simulations, where QCD is 'solved' by numerical Monte 
Carlo techniques on a discretized space-time lattice. Ideally, one would 
then use $\chi$PT to extract information on the LEC:s of QCD, as well as 
for the extrapolation of Lattice QCD results to the continuum limit. 
Unfortunately, $\chi$PT is insufficient for this purpose as Lattice QCD 
simulations have, for reasons of computational efficiency, to be performed 
in the so-called partially quenched approximation (PQQCD), where the masses 
of the sea quarks are different from and typically much larger than the 
valence quark masses. However, much progress is being made and at present, 
simulations are practicable for light sea quark masses of $\sim m_s/4$, 
where $m_s$ is the strange quark mass. Nevertheless, this is still quite 
far from the physical light quark masses of $\sim m_s/25$.

Partially Quenched $\chi$PT (PQ$\chi$PT) is the extension of $\chi$PT to 
the case of unequal valence and sea quark masses~\cite{QChPT,PQChPT}, and 
provides a way of making contact with present Lattice QCD simulations if 
the sea quark masses are small enough so that the chiral expansion is 
applicable. The possiblity of producing Lattice QCD data for different 
values of the sea and valence quark masses also allows for the 
investigation of a more comprehensive set of LEC:s. In PQ$\chi$PT, 
disconnected valence quark loops are eliminated by introduction of bosonic 
'ghost' quarks. PQ$\chi$PT at NLO as well as the treatment of the 
super-singlet $\Phi_0$ has been discussed in great detail 
in~\cite{SharpePQ}.

\section{PQ$\chi$PT at NNLO}

Since QCD may be obtained as a continuous limit of PQQCD, the LEC:s of 
PQQCD with $n_f$ sea quarks should, in general, be expected to be the same 
as for (unquenched) $n_f$ flavor QCD. A slight complication is introduced 
by the Cayley-Hamilton relations which are used to decrease the number of 
independent LEC:s in the unquenched case, since these no longer hold for 
partially quenched theories. Nevertheless, the LEC:s of unquenched QCD may 
be expressed as linear combinations of the PQQCD ones, so that a set of 
values for the partially quenched LEC:s may be directly translated into the 
physical ones.

\begin{figure}[h!]
  \includegraphics[width=\textwidth]{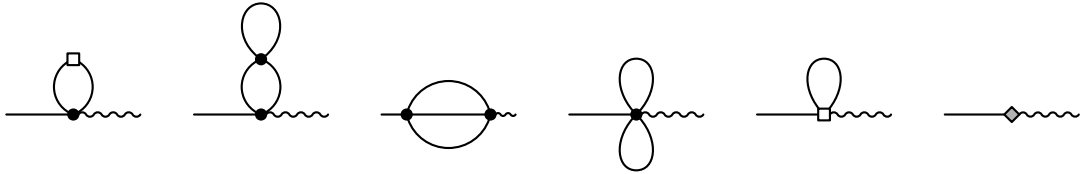}
  \caption{NNLO Feynman diagrams for the calculation of the pseudoscalar 
meson decay constant. Solid lines are pseudoscalars and wiggly lines are 
external axial currents. Solid dots, open squares and diamonds represent 
vertices of the $\mathcal{O}(p^2)$, $\mathcal{O}(p^4)$ and 
$\mathcal{O}(p^6)$ PQ$\chi$PT Lagrangians, respectively. Additional 
wavefunction renormalization contributions are not shown.}
  \label{feynfig}
\end{figure}

The expressions obtained at NNLO in PQ$\chi$PT are extremely long compared 
to the ones in unquenched $\chi$PT, even though the basic Feynman diagrams 
are similar. The NNLO Feynman diagrams for the calculation of the decay 
constants of the pseudoscalar mesons are shown in Fig.~\ref{feynfig}. One 
source of added complexity is the appearance of many more possible 
combinations of quark masses in the calculations, but most importantly, the 
meson propagators in the neutral sector of PQ$\chi$PT contain both single 
and double poles, the residues of which consist of complicated rational 
functions of the sea and valence quark masses. Fortunately, these residues 
satisfy a multitude of nontrivial relations which can, although at great 
effort, be used to compress the end results by significantly more than an 
order of magnitude, thereby bringing them to a publishable form. 

\begin{figure}[h!]
  \includegraphics[height=.5\textwidth]{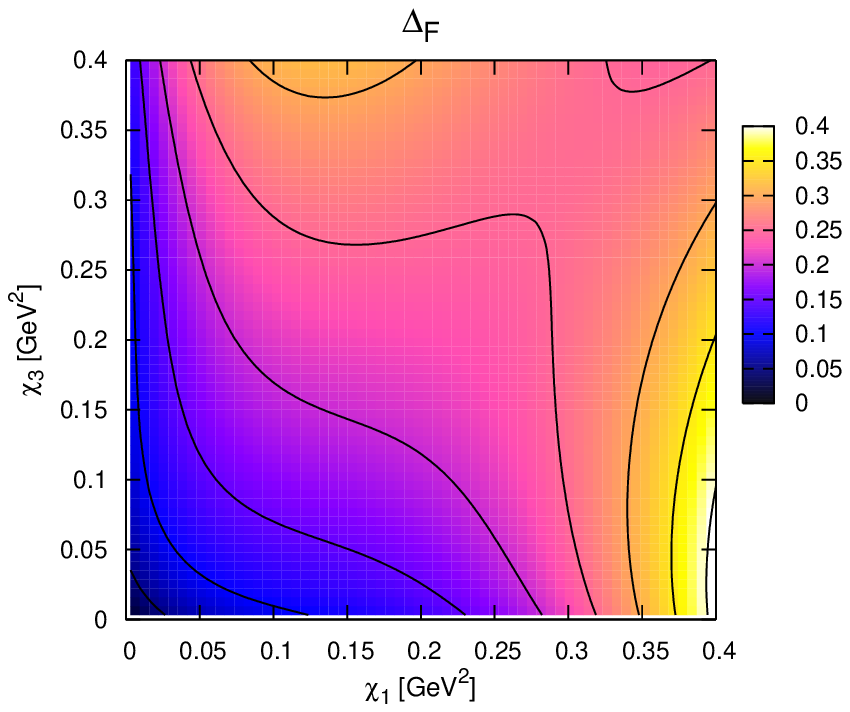}
  \includegraphics[height=.5\textwidth]{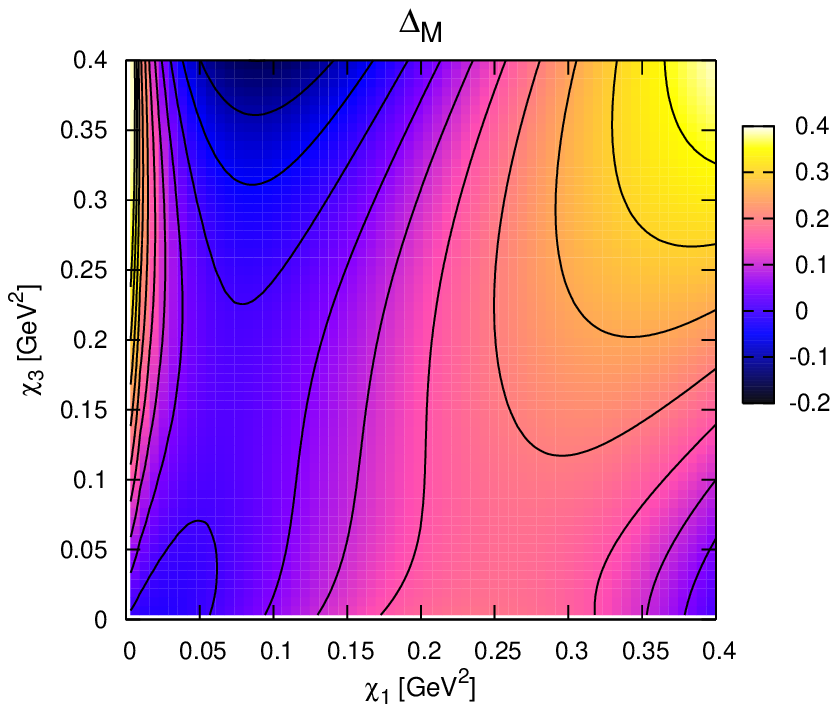}
  \caption{The relative deviation from the tree-level result in 
$n_f = 2$ PQ$\chi$PT at NNLO, for the pseudoscalar meson decay constant 
($\Delta_F$) and mass ($\Delta_M$), as a function of the valence ($m_1$) 
and sea quark ($m_3$) masses defined through $\chi_i \equiv 2B_0\,m_i$, for 
the set of LEC:s called 'fit 10' in ref.~\cite{fit10}.}
  \label{plotfig}
\end{figure}

The masses and decay constants at NNLO in PQ$\chi$PT depend on a number of 
as yet largely unknown LEC:s $K_i^r$ from the $\mathcal{O}(p^6)$ 
counterterms. As expected from earlier work at two loops in $\chi$PT, the 
contributions from the chiral logarithms are very large at NNLO, but are 
significantly moderated when realistic values of the $L_i^r$ are used. 
Numerical results up to NNLO for masses and decay constants for $n_f = 2$, 
with all $K_i^r$ set to zero, are shown in Fig.~\ref{plotfig}. The 
quantities plotted are $\Delta_F \equiv F\!/F_0 - 1$ for the decay 
constant, and $\Delta_M \equiv M^2\!/M_0^{\,2} - 1$ for the mass, where 
$F_0$ and $M_0^2$ denote the lowest order PQ$\chi$PT results.

\newpage

The following NNLO calculations in PQ$\chi$PT have already been published: 
The pseudoscalar meson mass for $n_f = 3$ sea quark flavors, for equal sea 
quark and equal valence quark masses~\cite{Lahde1}, the decay constant for 
$n_f = 3$ up to two nondegenerate sea quarks, and a complete calculation of 
masses and decay constants for $n_f = 2$~\cite{Lahde2}. The full results 
for $n_f = 3$ are being written up, along with a longer introduction to 
PQ$\chi$PT at NNLO.

\begin{theacknowledgments}
This work is supported by the European Union TMR network, Contract No. 
HPRN-CT-2002-00311 (EURIDICE) and by the European Community-Research 
Infrastructure Activity Contract No. RII3-CT-2004-506078 (HadronPhysics). 
TL acknowledges a travel grant from the Mikael Bj\"ornberg memorial 
foundation.
\end{theacknowledgments}

\bibliographystyle{aipproc}   

\end{document}